\documentclass[aps,prl,reprint,superscriptaddress,amssymb,amsmath]{revtex4-1}
\usepackage{graphicx}
\usepackage{dcolumn}
\usepackage{bm, color}
\usepackage[none]{hyphenat}

\begin{document}

\title{Strong anomalous Nernst effect in collinear magnetic Weyl semimetals without net magnetic moments}
\author{Jonathan Noky}
\affiliation{Max Planck Institute for Chemical Physics of Solids, D-01187 Dresden, Germany}
\author{Jacob Gayles}
\affiliation{Max Planck Institute for Chemical Physics of Solids, D-01187 Dresden, Germany}
\author{Claudia Felser} 
\affiliation{Max Planck Institute for Chemical Physics of Solids, D-01187 Dresden, Germany}
\author{Yan Sun} 
\email{ysun@cpfs.mpg.de}
\affiliation{Max Planck Institute for Chemical Physics of Solids, D-01187 Dresden, Germany}

\date{\today}

\begin{abstract}
We predict a large anomalous Nernst effect in the inverse Heusler compensated ferrimagnets
Ti$_2$Mn$X$ ($X$=Al,Ga,In) with vanishing net magnetic moments. Though the net magnetic
moment is zero, the Weyl points in these systems lead to a large anomalous Nernst conductivity (ANC) due to
the lack of \textcolor{black}{a global time-reversal symmetry operation} that \textcolor{black}{inverts} the sign of the Berry curvature.
In comparison to the noncollinear antiferromagnets Mn$_3$Sn and Mn$_3$Ge, the high ANC stems almost
entirely from the Weyl points in this class of compounds, and thus, it is topologically
protected. This work shows for the first time a large ANC with zero net magnetic moments in
collinear systems, which is helpful for comprehensive understanding of the thermoelectric
effect in zero-moment magnetic materials and its further applications.
\end{abstract}
\maketitle

In the ordinary Hall effect, a longitudinal electron current
generates a transverse voltage drop by the Lorentz force in
the presence of external magnetic fields. Contrary to this, the transverse electron current in the anomalous Hall
effect (AHE) is induced by intrinsic magnetic moments
and spin-orbit coupling (SOC)~\cite{Pugh_1953,Nagaosa_2010}.  
It is also possible to apply a temperature gradient, instead of an electric field, in combination with a magnetic field to 
generate a transverse charge current, which is known as the 
Nernst effect~\cite{Nerst_1887,WBauer2012}. Analogously,
a temperature-gradient-induced transverse charge current
can also exist in the absence of external magnetic fields, referred to as the anomalous Nernst effect (ANE)~\cite{WBauer2012,Lee_2004, XiaoDi_2006}. 
In ferromagnets, the imbalance of carriers with spin-up and spin-down leads to
a spin-polarized transverse charge current. Therefore,
the magnitude of the AHE and ANE \textcolor{black}{were historically considered} to be proportional to the
magnitude of the magnetic moments in the system~\cite{Nagaosa_2010}.

In the last decade, a more fundamental understanding 
of the intrinsic AHE from the Berry phase has been established~\cite{Xiao2010,Nagaosa_2010}. 
Because the Berry curvature (BC) is odd with respect to the
time reversal operation, the intrinsic AHE can only exist in magnetic systems. In collinear antiferromagnets (AFMs),
though the time reversal symmetry is broken, the combination of
time reversal and some space group operation is also a
symmetry of the system, which changes the sign of the BC~\cite{Zhang2014,Smejkal}. As a
consequence, the intrinsic anomalous Hall conductivity (AHC) obtained
from the integration of the BC in the whole Brillouin zone
is zero in collinear AFMs. The absence of such space group operations allows for the possibility of a large AHE in the noncollinear AFM Mn$_3$Ir~\cite{Chen_2014}. This lead to the inspiration to study the AHE and ANE in noncollinear AFMs with zero net magnetic moment due to 
the possible applications in spintronics. Furthermore, both strong AHE and 
ANE were observed in noncollinear AFM Mn$_3$Sn~\cite{Kubler_2014,Nakatsuji2015,Ikhlas_2017,Li_2017} 
and Mn$_3$Ge~\cite{Kubler_2014,Kiyohara_2016,Nayak2016} soon afterwards.

In compensated ferrimagnets (FiMs) with zero net magnetic
moment, owing to the absence of a symmetry operation that inverses the
sign of the BC, the AHE and ANE are
also allowed. Commonly in this type of systems the AHE is weak and cumbersome to detect due to the relatively low charge carrier density. However, if a compensated FiM possesses a special electronic band 
structure with a large BC, a strong AHE is expected. 
A typical example is the compensated ferrimagnetic Weyl semimetal 
(WSM)~\cite{Wan2011,Burkov:2011de}, in which the Weyl points behave as the monopoles of the
BC. 
\textcolor{black}{This leads to a large BC near the Weyl points and therefore to a large AHC, which is just the integration of the BC in the whole Brillouin zone.}
Based on this guiding principle, a strong AHE was 
recently predicted in the compensated ferrimagnetic Heusler WSM 
Ti$_2$Mn$X$ ($X$=Al, Ga, and In)~\cite{Shi_2018}. Due to the similarities of the underlying mechanisms, a strong ANE is also expected~\cite{XiaoDi_2006}. \textcolor{black}{However, the AHE is due to the BC of all occupied bands, whereas the ANE has contributions from both occupied and unoccupied bands near the Fermi level, which leads to two distinct behaviors of these effects.}

In this work, we have theoretically studied the ANE in the
compensated ferrimagnetic WSM Ti$_2$Mn$X$ ($X$=Al, Ga, and In) and complemented our results with a minimal model.
We predict that a large anomalous Nernst conductivity (ANC) can exist over a large temperature range. These results which indicate a
strong ANE in spite of a vanishing net magnetic moment are
especially interesting because, in particular, Ti$_2$MnAl has been
successfully grown in thin films where it shows a rather high
Curie temperature above $650\:\text{K}$~\cite{Feng2015}.

In this investigation we performed DFT calculations using the \textsc{vasp} package~\cite{kresse1996}.
We employed a plane wave basis set with pseudopotentials and used the generalized gradient approximation (GGA)~\cite{perdew1996} for the treatment of the exchange-correlation energy.
From the DFT band structure, Wannier functions were generated using
\textsc{wannier90}~\cite{Mostofi2008} with initial
projections to the s-, p-, and d-orbitals of Ti and Mn and to s-
and p-orbitals of $X$.
To evaluate the BC $\Omega$, the tight-binding Hamiltonian $H$ was set up from the Wannier functions and 
used with the Kubo formula~\cite{Thouless_1982,Xiao2010,Nagaosa_2010}
\begin{equation}
  \Omega_{ij}^n=\sum_{m\ne n} \frac{\langle n|\frac{\partial H}{\partial k_i}|m\rangle \langle m|\frac{\partial H}{\partial k_j}|n\rangle - (i \leftrightarrow j)}{(E_n-E_m)^2},
\end{equation}
where $\Omega_{ij}^n$ denotes the $ij$ component of the BC of the $n$-th band, $|n\rangle$ and $|m\rangle$ are the eigenstates of $H$, and $E_n$ and $E_m$ are the corresponding eigenvalues.
From this we calculate the $ij$ component of the AHC $\sigma_{ij}$ as
\begin{equation}
 \sigma_{ij}=\frac{e^2}{\hbar} \sum_n^{occ}\int \frac{d^3k}{(2\pi)^3}\Omega_{ij}^n
\end{equation}y
and of the ANC $\alpha_{ij}$ as proposed by Xiao et al. \cite{Xiao2010, XiaoDi_2006}
\begin{align}
  \label{eq:anc}
  \alpha_{ij}=\frac{1}{T} \frac{e}{\hbar} \sum_n \int \frac{d^3k}{(2\pi)^3} \Omega_{ij}^{n}[(E_{n}-E_F)f_{n}+\nonumber \\
  +k_BT\ln{(1+\exp{\frac{E_{n}-E_F}{-k_BT}})}],
\end{align}
where $T$ is the actual temperature, $f_{n}$ is the Fermi distribution, 
and $E_F$ is the Fermi level. To realize integrations over the Brillouin 
zone, a k mesh of $251\times 251\times 251$ points was used.


\begin{figure}[htb]
\centering
\includegraphics[width=0.48\textwidth]{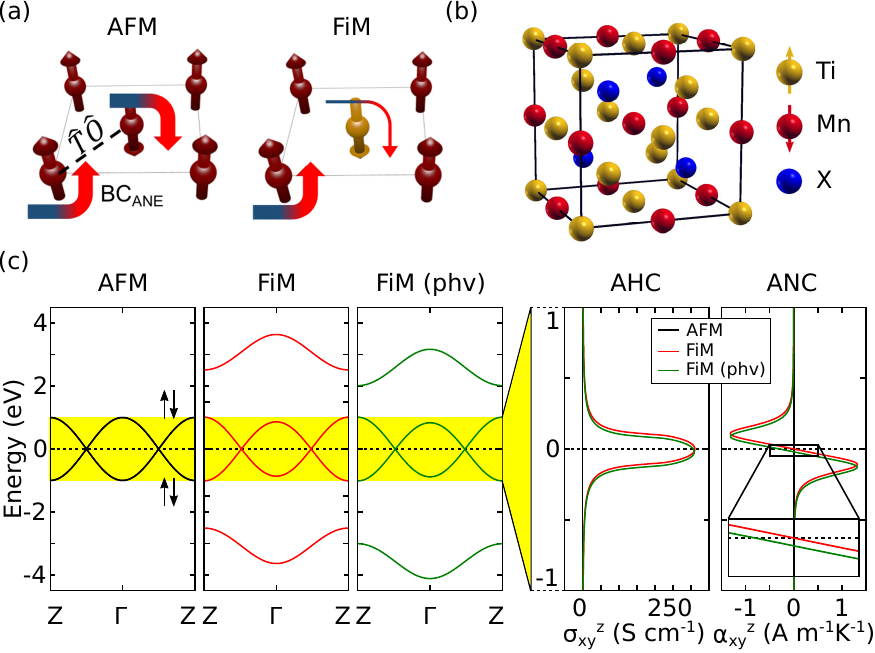}
   \caption{
   (a) Left: AFM structure with a time reversal + slide symmetry leading to a vanishing AHE. Right: Ferrimagnet with broken slide symmetry. (b) Inverted Heusler FCC crystal structure of Ti$_2$Mn$X$ ($X$=Al,Mn,Ga). 
The magnetic moments of Ti and Mn are all aligned along the (001) direction and 
compensate each other. (c) Band structure, AHC, and ANC (at $300$ K) of the model Hamiltonian for the AFM, FiM, and FiM (phv) case (for parameters see main text). The FiM systems show a non-zero AHC, but to get a finite ANC additionally the particle-hole symmetry has to be broken (see inset of the ANC). Note the change of the energy scale for AHC and ANC.}
\label{fig:model}
\end{figure}

As previously mentioned, the net Berry phase of an AFM is zero due
to the combined symmetry of a space group operation $\hat{O}$ and the
time reversal operation $\hat{T}$, that changes the sign of the Berry
curvature (BC). As an example, this can be observed in the combination symmetry $\hat{T}\hat{O}$ of a glide operation
to the center of the unit cell and time reversal, see the left panel in Fig. \ref{fig:model}(a). A simple and effective way to remove this symmetry
is by replacing the equivalent atoms lying on the other
sublattice with a different element, see the right panel in Fig. \ref{fig:model}(a), which is
just a compensated FiM, and a nonzero Berry
phase from the whole BZ is allowed. In both panels the flat arrows (BC$_\text{ANE}$) depict the electron flow from cold (blue) to hot (red) in the ANE. The FiM model is deduced from the compensated FiM Ti$_2$Mn$X$ ($X$=Al, Ga, In) [see Fig.~\ref{fig:model}(b)]. Since the charge
carrier density is relatively small in most compensated
FiMs, the net Berry phases are normally very
close to zero. However, the Berry phase can be strongly
enhanced by some topologically protected band structures, such as nodal
lines and Weyl points [see Fig. \ref{fig:model}(c)].

To achieve a deeper understanding of the underlying mechanisms, we first study a minimal effective four-band model derived from the two-band model by Lu~\textit{et al.}~\cite{Lu_2015} to analyze the effect of Weyl points on the ANE. The Hamiltonian reads as
\begin{align}
 H&=M_0\tau_z\otimes\sigma_z+\alpha(\sin{k_x}\tilde{\tau}\otimes\sigma_x+\sin{k_y}\tilde{\tau}\otimes\sigma_y) \nonumber\\
 &+tC(\vec{k})\tilde{\tau}\otimes\sigma_z+t'C(\vec{k})\tau_x\otimes\sigma_x\\
 &\text{with }C(\vec{k})=\cos{k_x}+\cos{k_y}+\cos{k_z}, \nonumber
\end{align}
where $\tau$ and $\sigma$ are the Pauli matrices for lattice and spin, respectively, and $M_0$, $\alpha$, $t$, and $t'$ are model parameters. The matrix $\tilde{\tau}$ is used to switch between AFM and FiM and is defined below. This minimal model describes a pair of Dirac/Weyl nodes at $k_x=k_y=0$ with all their topological properties. As model parameters we used $M_0=2\:\text{eV}$, $\alpha=0.1\:\text{eV}$, $t=-1\:\text{eV}$, and $t'=0.2\:\text{eV}$ and the results discussed in the following are all shown in Fig.~\ref{fig:model}(c). To investigate the AFM case (see Fig.~\ref{fig:model}(a) left panel) we set $\tilde{\tau}=\tau_z$ and $t'=0$, which leads to a band structure with two Dirac points and vanishing AHC and ANC due to the presence of the $\hat{T}\hat{O}$ symmetry. To break this symmetry we make a transition to the FiM case (see Fig.~\ref{fig:model}(a) right panel) by setting $\tilde{\tau}=\begin{pmatrix} 1 & 0 \\ 0 & 0.5\end{pmatrix}$ to model different atom types at the lattice sites. We also set $t'=0.2\:\text{eV}$ to switch on hopping between the sites. The resulting band structure exhibits two Weyl points at the Fermi level and a large AHC connected to them. However, the ANC at the Fermi level is still zero due to the preserved particle-hole symmetry (PHS) in the band structure. Therefore, it is crucial to get to a PHS violation (phv) to find a finite ANC. In our model we achieve this by shifting the onsite energy of the last two orbitals up by $1\:\text{eV}$ simulating the difference between the atom types. Just like in the symmetry preserving case, the model shows two Weyl points on the line $k_x=k_y=0$. However, since 
the last term now breaks the PHS, the band structure and thus the 
AHC are no longer symmetric around the Fermi level, leading to a nonzero ANC at zero energy.
In both FiM cases the ANC vanishes when the AHC reaches its 
maximum value.
It is important to note, that SOC (model parameter $\alpha$) is necessary to get a finite AHC. \textcolor{black}{However, the strength of the AHC mainly is proportional to the distance of the two Weyl points in k space \cite{Armitage2018} and shows to be inpendent of $\alpha$, when SOC is large.}
This minimal model captures all important properties regarding Weyl points, AHE, and ANE, and motivates us to study the compensated FiMs Ti$_2$Mn$X$ ($X$=Al, Ga, In).

\begin{figure}[htb]
\centering
\includegraphics[width=0.48\textwidth]{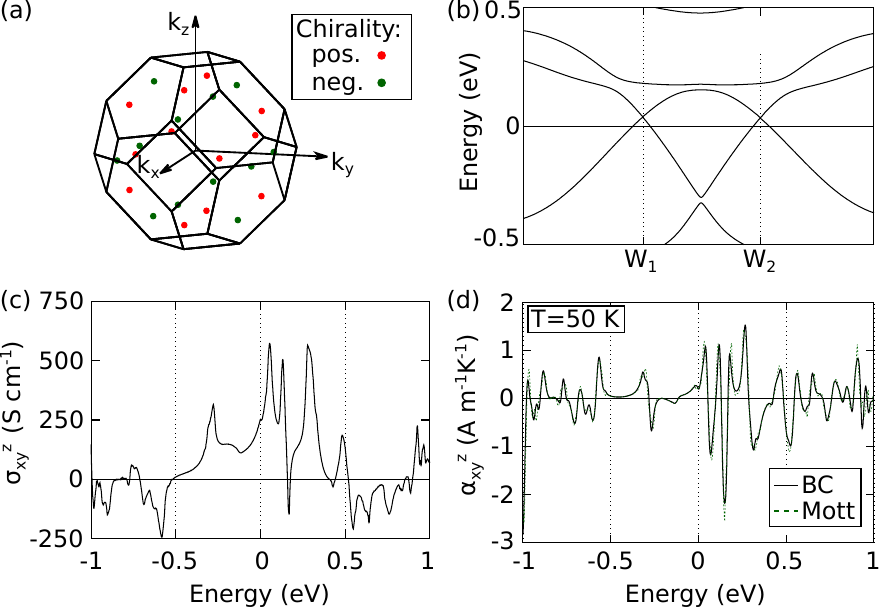}
   \caption{(a) Brillouin zone of Ti$_2$MnAl with the location of the 
12 pairs of Weyl points located $34-40\:\text{meV}$ above the Fermi level. (b) Band structure of 
Ti$_2$MnAl through two Weyl points at W$_1$ and W$_2$. (c) AHC for Ti$_2$MnAl. A maximum linked 
to the Weyl points can be seen $\approx 40\:\text{meV}$ above the Fermi level. (d) ANC in Ti$_2$MnAl at $50\:\text{K}$ calculated using the BC formalism and the Mott relation.
}
\label{fig:material}
\end{figure}

These compounds have an inverse Heusler lattice structure with space
group $F\bar{4}3m$ (No. 216) (see Fig. \ref{fig:model}(b))~\cite{Feng2015}.
They exhibit an isotropic ferrimagnetic structure, where magnetic moments are located
at the Ti [$\mu\approx1.3(1.2)\:\mu_B$ for first(second) atom] and Mn ($\mu\approx-2.5\:\mu_B$) atoms. The net magnetic moment in Ti$_2$Mn$X$ vanishes because of the compensated magnetic sublattices formed by Ti and Mn.
In total, there are twelve pairs of Weyl points. Their positions in the Brillouin zone are depicted in
Fig. \ref{fig:material}(a) and they are located slightly above
the Fermi level [$34-40(27-36,13-27)\:\text{meV}$ for Ti$_2$MnAl(Ga,In), respectively],
as indicated in Fig. \ref{fig:material}(b). The influence of the SOC becomes evident: While in its absence all Weyl points are at the same energy, the different SOC strength [$\Delta_{SOC} = 0.2(5.5,40.2)\:\text{meV}$ for Al(Ga,In)] leads to a larger spread in energy as the atoms get heavier.

At low temperature, the ANC can be obtained from the
Mott relation as the derivative of AHC with respect to energy~\cite{XiaoDi_2006, Xiao2010},
\begin{equation}
  \label{eq:mott}
  \alpha_{ij} =  \frac{\pi^2}{3} \frac{k_B^2 T}{e} \frac{\partial\sigma_{ij}}{\partial E}(E_F).
\end{equation}

To determine the ANE at low temperature, we first calculated the energy-dependent AHC for the
three compounds with moments aligned along the (001) direction. As shown in Fig. \ref{fig:material}(c) for Ti$_2$MnAl,
the AHC ($\sigma_{xy}^z$) can reach up to $253(268,133)\:\text{S cm}^{-1}$ for Ti$_2$MnAl(Ga,In), respectively.
The AHC varies sharply in the energy space around the Fermi level, which leads to a
large ANC ($\alpha_{xy}^z$) [see Fig. \ref{fig:material}(d)]. At $50\:\text{K}$, the ANCs for Ti$_2$MnAl(Ga,In)
are approximately $0.2(0.16,-0.11)\:\text{A m}^{-1}\text{K}^{-1}$ as large as that in the noncollinear AFM Mn$_3$Sn~\cite{Ikhlas_2017,Li_2017}. 
The ANC from the BC formalism in equation (\ref{eq:anc}) and from the Mott relation in equation (\ref{eq:mott}) are compared in Fig. \ref{fig:material}(d) at $50\:\text{K}$ for Ti$_2$MnAl and show a very good agreement.

\begin{figure}[htb]
\centering
\includegraphics[width=0.48\textwidth]{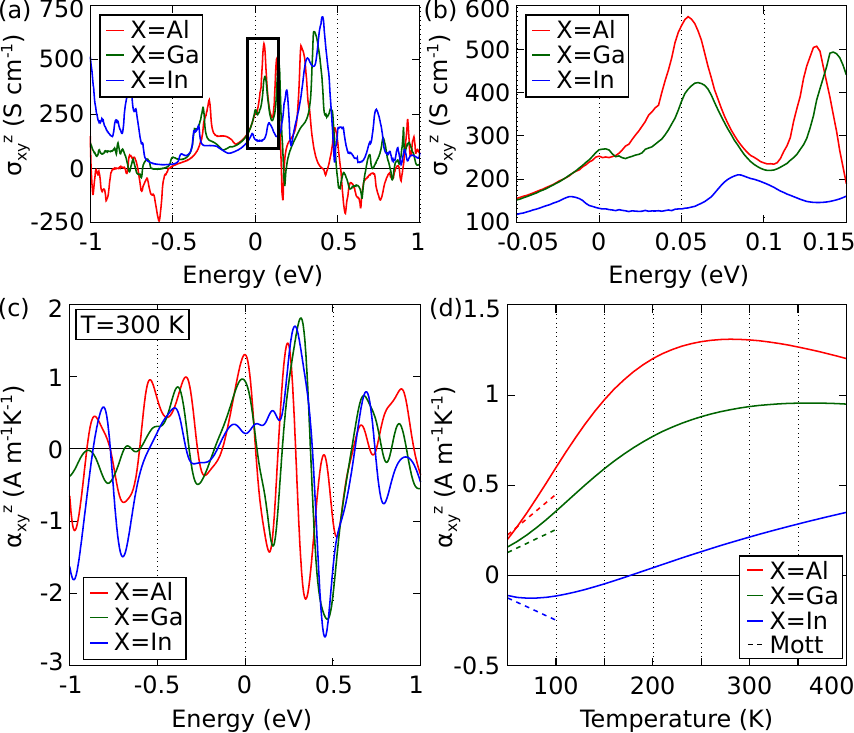}
   \caption{(a) The dependence of the AHC of Ti$_2$Mn$X$ ($X$=Al,Ga,In) on energy. The data in the black rectangle is shown enlarged in (b). (c) The dependence of the ANC of Ti$_2$Mn$X$ ($X$=Al,Ga,In) on energy at
$T=300\:\text{K}$. (d) The dependence of the ANC of the three components at $E=E_F$ on temperature. The dashed lines between $50\:\text{K}$ and $100\:\text{K}$ show
the low-energy approximation (Mott relation).
}
\label{fig:anct}
\end{figure}

From an application point of view, the ANC at room temperature is interesting.
In Fig. \ref{fig:anct}(c) we see the energy dependence of $\alpha_{xy}^{z}$ at $300\:\text{K}$.
At the Fermi energy, the value shows a peak of $\alpha_{xy}^z = 1.31(0.94)\:\text{A m}^{-1}\text{K}^{-1}$ for Ti$_2$MnAl(Ga), which is a high value in comparison to Mn$_3$Sn~\cite{Ikhlas_2017,Li_2017}. In an analogous manner to the minimal model established in the beginning, the ANC also drops to
zero at an energy above the Fermi level, which coincides with a maximum of the AHC.
The different behavior of Ti$_2$MnIn can be understood in terms of the stronger SOC induced by the higher atomic mass of In. This leads to a larger spread in energy of the Weyl points and strongly influences the behavior of the AHC [see Fig. \ref{fig:anct} (a) and (b)], and consequently, the ANC near the Fermi level. Based on this concept, this indicates a strong connection between the Weyl points and AHE/ANE in these systems. The sign of the ANC is related to the slope of the AHC at the Fermi level via the Mott relation. This slope is positive for Ti$_2$MnAl(Ga) and negative for Ti$_2$MnIn, which is also the case for the respective ANCs at low temperature.

We also investigated the temperature dependence of the ANC [see Fig. \ref{fig:anct}(d)] using equation (\ref{eq:anc}).
The temperature dependent ANC from the Mott relation and equation (\ref{eq:anc})
shows good agreement at low temperatures. The divergence between the two at
high temperatures imply the Mott relation only applies to low temperatures.
The ANC shows a broad peak around $T = 300(350)\:\text{K}$ for Ti$_2$MnAl(Ga). This high
value in this temperature range means that both Ti$_2$MnAl and Ti$_2$MnGa are interesting candidate
materials for room-temperature thermoelectric applications. Ti$_2$MnIn behaves differently due to a larger SOC. More importantly, the sign of the ANC changes near $175\:\text{K}$.

\begin{figure}
\centering
\includegraphics[width=0.48\textwidth]{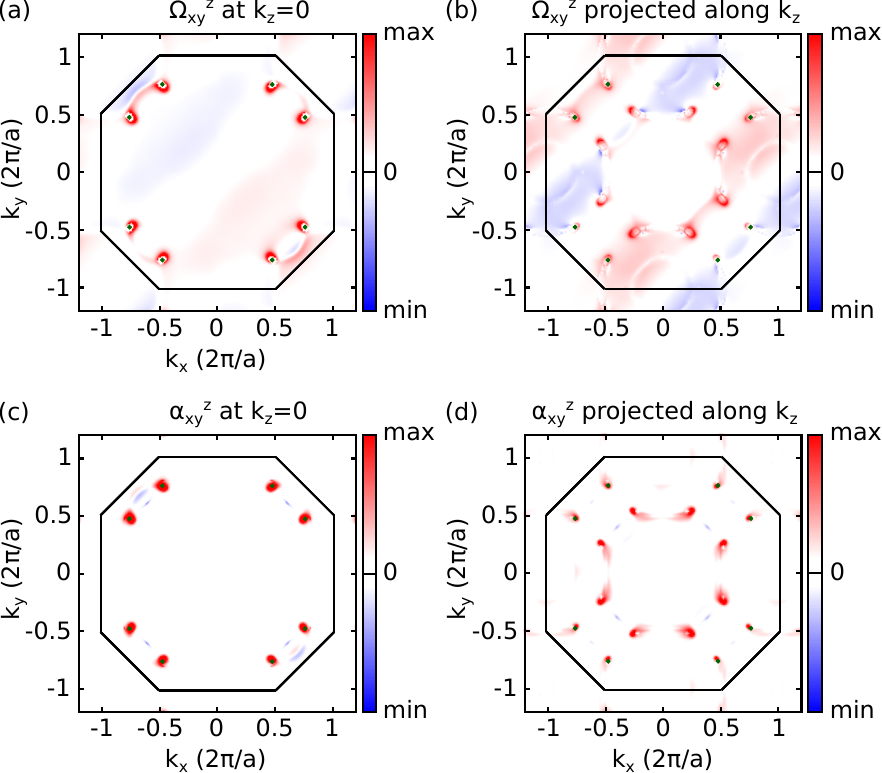}
   \caption{(a) BC in the $k_x$-$k_y$ plane at $k_z=0$, showing only contributions near the Weyl points (marked as green dots). (b) \textcolor{black}{BC integrated along $k_z$ ($\Omega_{xy}^{z,int}(k_x,k_y)=\int{\Omega_{xy}^{z}(k_x,k_y,k_z)dk_z}$)}, again the highest contributions come from the Weyl points. The eight more central maxima can be associated to the same Weyl points but from the neighboring BZs and are included due to the projection procedure.
(c) ANC at $300\:\text{K}$ in the $k_x$-$k_y$ plane at $k_z=0$, showing only contributions near the Weyl points. (d) ANC at $300\:\text{K}$ integrated along $k_z$. A clear connection to the BC in (b) is visible.}
\label{fig:bc}
\end{figure}

Since the ANC can be understood as the integration of the BC over
the BZ with the inclusion of temperature effects, we further examined the origin of
the ANC by investigating the BC distributions in BZ.
Structurally and energetically all three compounds share similar properties, therefore Ti$_2$MnAl with
magnetization along the (001) direction is suitable for the $\textbf{k}$ dependent BC analysis.
In Fig. \ref{fig:bc}(a) the $k_x$-$k_y$-plane at $k_z = 0$ is shown.
There are eight spots with a non-negligible $\Omega_{xy}^z$ component.
These points correspond to the positions of the Weyl points near $k_z = 0$
(marked as green dots). Examining the \textcolor{black}{BC integrated
along $k_z$ ($\Omega_{xy}^{z,int}(k_x,k_y)=\int{\Omega_{xy}^{z}(k_x,k_y,k_z)dk_z}$)}, again there are clear maximums located at the projected positions
of the Weyl points. The eight points with a high BC contribution
which are closer to the center of the Brillouin zone can be referred to the
same eight Weyl points but in the neighboring Brillouin zones, and result
from the projection.

Taking temperature into consideration, as indicated in Fig. \ref{fig:bc}(c) a cut through the Brillouin zone at $k_z=0$ is shown with the z-component
of the ANC at $300\:\text{K}$. Moreover, the highest contributions to the ANC stem from the eight
Weyl points near the plane in this case. Examining the projection of the ANC along $k_z$
[see Fig. \ref{fig:bc}(d)] a clear relation to the BC in
Fig. \ref{fig:bc}(b) is visible. Thus, the ANC stems mostly from the Weyl points
in these materials.

In summary, we have theoretically investigated the ANE in the compensated ferrimagnetic
WSMs Ti$_2$Mn$X$ ($X$=Al,Ga,In). Although the net magnetic moments are zero, all the three
compounds exhibit strong ANEs due to the large BC of the Weyl points around the Fermi level. 
In comparison to the noncollinear AFMs Mn$_3$Sn and Mn$_3$Ge, which also exhibit nonzero 
ANC in the absence of net magnetic moments, the ANC in Ti$_2$Mn$X$ ($X$=Al,Ga,In)
is dominated by the Weyl points. 
The temperature-dependent ANC shows a broad plateau around $T=300\:\text{K}$ in the compounds Ti$_2$MnAl and Ti$_2$MnGa.
Due to the large ANC and the vanishing net magnetic moment, they would allow for possible application and detection at ambient temperature. In studies of topological bands in AFM systems it has been shown that there is an anisotropic response with the direction of the magnetization, which we believe should also be present in our systems due to the similarities~\cite{Smejkal,Chadov2017}. This work demonstrates, for the first time, a large ANC in fully compensated collinear ferrimagnetic systems that is due to the topology of the electronic structure.

This work was financially supported by the ERC
Advanced Grant No.
291472 ``Idea Heusler'', ERC
Advanced Grant No. 742068 ``TOPMAT'', and ``ASPIN'' (EU H2020 FET Open Grant No. 766566).


\begin{thebibliography}{27}%
\makeatletter
\providecommand \@ifxundefined [1]{%
 \@ifx{#1\undefined}
}%
\providecommand \@ifnum [1]{%
 \ifnum #1\expandafter \@firstoftwo
 \else \expandafter \@secondoftwo
 \fi
}%
\providecommand \@ifx [1]{%
 \ifx #1\expandafter \@firstoftwo
 \else \expandafter \@secondoftwo
 \fi
}%
\providecommand \natexlab [1]{#1}%
\providecommand \enquote  [1]{``#1''}%
\providecommand \bibnamefont  [1]{#1}%
\providecommand \bibfnamefont [1]{#1}%
\providecommand \citenamefont [1]{#1}%
\providecommand \href@noop [0]{\@secondoftwo}%
\providecommand \href [0]{\begingroup \@sanitize@url \@href}%
\providecommand \@href[1]{\@@startlink{#1}\@@href}%
\providecommand \@@href[1]{\endgroup#1\@@endlink}%
\providecommand \@sanitize@url [0]{\catcode `\\12\catcode `\$12\catcode
  `\&12\catcode `\#12\catcode `\^12\catcode `\_12\catcode `\%12\relax}%
\providecommand \@@startlink[1]{}%
\providecommand \@@endlink[0]{}%
\providecommand \url  [0]{\begingroup\@sanitize@url \@url }%
\providecommand \@url [1]{\endgroup\@href {#1}{\urlprefix }}%
\providecommand \urlprefix  [0]{URL }%
\providecommand \Eprint [0]{\href }%
\providecommand \doibase [0]{http://dx.doi.org/}%
\providecommand \selectlanguage [0]{\@gobble}%
\providecommand \bibinfo  [0]{\@secondoftwo}%
\providecommand \bibfield  [0]{\@secondoftwo}%
\providecommand \translation [1]{[#1]}%
\providecommand \BibitemOpen [0]{}%
\providecommand \bibitemStop [0]{}%
\providecommand \bibitemNoStop [0]{.\EOS\space}%
\providecommand \EOS [0]{\spacefactor3000\relax}%
\providecommand \BibitemShut  [1]{\csname bibitem#1\endcsname}%
\let\auto@bib@innerbib\@empty
\bibitem [{\citenamefont {Pugh}\ and\ \citenamefont
  {Rostoker}(1953)}]{Pugh_1953}%
  \BibitemOpen
  \bibfield  {author} {\bibinfo {author} {\bibfnamefont {E.~M.}\ \bibnamefont
  {Pugh}}\ and\ \bibinfo {author} {\bibfnamefont {N.}~\bibnamefont
  {Rostoker}},\ }\href@noop {} {\bibfield  {journal} {\bibinfo  {journal} {Rev.
  Mod. Phys.}\ }\textbf {\bibinfo {volume} {25}},\ \bibinfo {pages} {151}
  (\bibinfo {year} {1953})}\BibitemShut {NoStop}%
\bibitem [{\citenamefont {Nagaosa}\ \emph {et~al.}(2010)\citenamefont
  {Nagaosa}, \citenamefont {Sinova}, \citenamefont {Onoda}, \citenamefont
  {MacDonald},\ and\ \citenamefont {Ong}}]{Nagaosa_2010}%
  \BibitemOpen
  \bibfield  {author} {\bibinfo {author} {\bibfnamefont {N.}~\bibnamefont
  {Nagaosa}}, \bibinfo {author} {\bibfnamefont {J.}~\bibnamefont {Sinova}},
  \bibinfo {author} {\bibfnamefont {S.}~\bibnamefont {Onoda}}, \bibinfo
  {author} {\bibfnamefont {A.~H.}\ \bibnamefont {MacDonald}}, \ and\ \bibinfo
  {author} {\bibfnamefont {N.~P.}\ \bibnamefont {Ong}},\ }\href@noop {}
  {\bibfield  {journal} {\bibinfo  {journal} {Rev. Mod. Phys.}\ }\textbf
  {\bibinfo {volume} {82}},\ \bibinfo {pages} {1539} (\bibinfo {year}
  {2010})}\BibitemShut {NoStop}%
\bibitem [{\citenamefont {Nernst}(1887)}]{Nerst_1887}%
  \BibitemOpen
  \bibfield  {author} {\bibinfo {author} {\bibfnamefont {W.}~\bibnamefont
  {Nernst}},\ }\href@noop {} {\bibfield  {journal} {\bibinfo  {journal} {Ann.
  Phys.}\ }\textbf {\bibinfo {volume} {267}},\ \bibinfo {pages} {760} (\bibinfo
  {year} {1887})}\BibitemShut {NoStop}%
\bibitem [{\citenamefont {{W Bauer}}\ \emph {et~al.}(2012)\citenamefont {{W
  Bauer}}, \citenamefont {Saitoh},\ and\ \citenamefont {van
  Wees}}]{WBauer2012}%
  \BibitemOpen
  \bibfield  {author} {\bibinfo {author} {\bibfnamefont {G.~E.}\ \bibnamefont
  {{W Bauer}}}, \bibinfo {author} {\bibfnamefont {E.}~\bibnamefont {Saitoh}}, \
  and\ \bibinfo {author} {\bibfnamefont {B.~J.}\ \bibnamefont {van Wees}},\
  }\href {https://www.nature.com/articles/nmat3301.pdf} {\bibfield  {journal}
  {\bibinfo  {journal} {Nat. Mat.}\ }\textbf {\bibinfo {volume} {11}},\
  \bibinfo {pages} {391} (\bibinfo {year} {2012})}\BibitemShut {NoStop}%
\bibitem [{\citenamefont {Lee}\ \emph {et~al.}(2004)\citenamefont {Lee},
  \citenamefont {Watauchi}, \citenamefont {Miller}, \citenamefont {Cava},\ and\
  \citenamefont {Ong}}]{Lee_2004}%
  \BibitemOpen
  \bibfield  {author} {\bibinfo {author} {\bibfnamefont {W.-L.}\ \bibnamefont
  {Lee}}, \bibinfo {author} {\bibfnamefont {S.}~\bibnamefont {Watauchi}},
  \bibinfo {author} {\bibfnamefont {V.~L.}\ \bibnamefont {Miller}}, \bibinfo
  {author} {\bibfnamefont {R.~J.}\ \bibnamefont {Cava}}, \ and\ \bibinfo
  {author} {\bibfnamefont {N.~P.}\ \bibnamefont {Ong}},\ }\href@noop {}
  {\bibfield  {journal} {\bibinfo  {journal} {Phys. Rev. Lett.}\ }\textbf
  {\bibinfo {volume} {93}},\ \bibinfo {pages} {226601} (\bibinfo {year}
  {2004})}\BibitemShut {NoStop}%
\bibitem [{\citenamefont {Xiao}\ \emph {et~al.}(2006)\citenamefont {Xiao},
  \citenamefont {Yao}, \citenamefont {Fang},\ and\ \citenamefont
  {Niu}}]{XiaoDi_2006}%
  \BibitemOpen
  \bibfield  {author} {\bibinfo {author} {\bibfnamefont {D.~X.}\ \bibnamefont
  {Xiao}}, \bibinfo {author} {\bibfnamefont {Y.}~\bibnamefont {Yao}}, \bibinfo
  {author} {\bibfnamefont {Z.}~\bibnamefont {Fang}}, \ and\ \bibinfo {author}
  {\bibfnamefont {Q.}~\bibnamefont {Niu}},\ }\href@noop {} {\bibfield
  {journal} {\bibinfo  {journal} {Phys. Rev. Lett.}\ }\textbf {\bibinfo
  {volume} {97}},\ \bibinfo {pages} {026603} (\bibinfo {year}
  {2006})}\BibitemShut {NoStop}%
\bibitem [{\citenamefont {Xiao}\ \emph {et~al.}(2010)\citenamefont {Xiao},
  \citenamefont {Chang},\ and\ \citenamefont {Niu}}]{Xiao2010}%
  \BibitemOpen
  \bibfield  {author} {\bibinfo {author} {\bibfnamefont {D.}~\bibnamefont
  {Xiao}}, \bibinfo {author} {\bibfnamefont {M.-C.}\ \bibnamefont {Chang}}, \
  and\ \bibinfo {author} {\bibfnamefont {Q.}~\bibnamefont {Niu}},\ }\href@noop
  {} {\bibfield  {journal} {\bibinfo  {journal} {Rev. Mod. Phys.}\ }\textbf
  {\bibinfo {volume} {82}},\ \bibinfo {pages} {1959} (\bibinfo {year}
  {2010})}\BibitemShut {NoStop}%
\bibitem [{\citenamefont {Zhang}\ \emph {et~al.}(2014)\citenamefont {Zhang},
  \citenamefont {Liu}, \citenamefont {Luo}, \citenamefont {Freeman},\ and\
  \citenamefont {Zunger}}]{Zhang2014}%
  \BibitemOpen
  \bibfield  {author} {\bibinfo {author} {\bibfnamefont {X.}~\bibnamefont
  {Zhang}}, \bibinfo {author} {\bibfnamefont {Q.}~\bibnamefont {Liu}}, \bibinfo
  {author} {\bibfnamefont {J.~W.}\ \bibnamefont {Luo}}, \bibinfo {author}
  {\bibfnamefont {A.~J.}\ \bibnamefont {Freeman}}, \ and\ \bibinfo {author}
  {\bibfnamefont {A.}~\bibnamefont {Zunger}},\ }\href@noop {} {\bibfield
  {journal} {\bibinfo  {journal} {Nature Physics}\ } (\bibinfo {year}
  {2014})},\ \Eprint {http://arxiv.org/abs/1402.4446} {1402.4446} \BibitemShut
  {NoStop}%
\bibitem [{\citenamefont {{\v{S}}mejkal}\ \emph {et~al.}()\citenamefont
  {{\v{S}}mejkal}, \citenamefont {{\v{Z}}elezn{\'{y}}}, \citenamefont
  {Sinova},\ and\ \citenamefont {Jungwirth}}]{Smejkal}%
  \BibitemOpen
  \bibfield  {author} {\bibinfo {author} {\bibfnamefont {L.}~\bibnamefont
  {{\v{S}}mejkal}}, \bibinfo {author} {\bibfnamefont {J.}~\bibnamefont
  {{\v{Z}}elezn{\'{y}}}}, \bibinfo {author} {\bibfnamefont {J.}~\bibnamefont
  {Sinova}}, \ and\ \bibinfo {author} {\bibfnamefont {T.}~\bibnamefont
  {Jungwirth}},\ }\href
  {https://journals.aps.org/prl/pdf/10.1103/PhysRevLett.118.106402} {\
  }\BibitemShut {NoStop}%
\bibitem [{\citenamefont {Chen}\ \emph {et~al.}(2014)\citenamefont {Chen},
  \citenamefont {Niu},\ and\ \citenamefont {MacDonald}}]{Chen_2014}%
  \BibitemOpen
  \bibfield  {author} {\bibinfo {author} {\bibfnamefont {H.}~\bibnamefont
  {Chen}}, \bibinfo {author} {\bibfnamefont {Q.}~\bibnamefont {Niu}}, \ and\
  \bibinfo {author} {\bibfnamefont {A.~H.}\ \bibnamefont {MacDonald}},\
  }\href@noop {} {\bibfield  {journal} {\bibinfo  {journal} {Phys. Rev. Lett.}\
  }\textbf {\bibinfo {volume} {112}},\ \bibinfo {pages} {017205} (\bibinfo
  {year} {2014})}\BibitemShut {NoStop}%
\bibitem [{\citenamefont {Kuebler}\ and\ \citenamefont
  {Felser}(2014)}]{Kubler_2014}%
  \BibitemOpen
  \bibfield  {author} {\bibinfo {author} {\bibfnamefont {J.}~\bibnamefont
  {Kuebler}}\ and\ \bibinfo {author} {\bibfnamefont {C.~F.}\ \bibnamefont
  {Felser}},\ }\href@noop {} {\bibfield  {journal} {\bibinfo  {journal} {EPL}\
  }\textbf {\bibinfo {volume} {108}},\ \bibinfo {pages} {67001} (\bibinfo
  {year} {2014})}\BibitemShut {NoStop}%
\bibitem [{\citenamefont {Nakatsuji}\ \emph {et~al.}(2015)\citenamefont
  {Nakatsuji}, \citenamefont {Kiyohara},\ and\ \citenamefont
  {Higo}}]{Nakatsuji2015}%
  \BibitemOpen
  \bibfield  {author} {\bibinfo {author} {\bibfnamefont {S.}~\bibnamefont
  {Nakatsuji}}, \bibinfo {author} {\bibfnamefont {N.}~\bibnamefont {Kiyohara}},
  \ and\ \bibinfo {author} {\bibfnamefont {T.}~\bibnamefont {Higo}},\
  }\href@noop {} {\bibfield  {journal} {\bibinfo  {journal} {Nature}\ }\textbf
  {\bibinfo {volume} {527}},\ \bibinfo {pages} {212} (\bibinfo {year}
  {2015})}\BibitemShut {NoStop}%
\bibitem [{\citenamefont {Ikhlas}\ \emph {et~al.}(2017)\citenamefont {Ikhlas},
  \citenamefont {Tomita}, \citenamefont {Koretsune}, \citenamefont {Suzuki},
  \citenamefont {Nishio-Hamane}, \citenamefont {Arita}, \citenamefont {Otani},\
  and\ \citenamefont {Nakatsuji}}]{Ikhlas_2017}%
  \BibitemOpen
  \bibfield  {author} {\bibinfo {author} {\bibfnamefont {M.}~\bibnamefont
  {Ikhlas}}, \bibinfo {author} {\bibfnamefont {T.}~\bibnamefont {Tomita}},
  \bibinfo {author} {\bibfnamefont {T.}~\bibnamefont {Koretsune}}, \bibinfo
  {author} {\bibfnamefont {M.-T.}\ \bibnamefont {Suzuki}}, \bibinfo {author}
  {\bibfnamefont {D.}~\bibnamefont {Nishio-Hamane}}, \bibinfo {author}
  {\bibfnamefont {R.}~\bibnamefont {Arita}}, \bibinfo {author} {\bibfnamefont
  {Y.}~\bibnamefont {Otani}}, \ and\ \bibinfo {author} {\bibfnamefont
  {S.}~\bibnamefont {Nakatsuji}},\ }\href@noop {} {\bibfield  {journal}
  {\bibinfo  {journal} {Nat. Phys.}\ }\textbf {\bibinfo {volume} {13}},\
  \bibinfo {pages} {1085} (\bibinfo {year} {2017})}\BibitemShut {NoStop}%
\bibitem [{\citenamefont {Li}\ \emph {et~al.}(2017)\citenamefont {Li},
  \citenamefont {Xu}, \citenamefont {Ding}, \citenamefont {Wang}, \citenamefont
  {Shen}, \citenamefont {Lu}, \citenamefont {Zhu},\ and\ \citenamefont
  {Behnia}}]{Li_2017}%
  \BibitemOpen
  \bibfield  {author} {\bibinfo {author} {\bibfnamefont {X.}~\bibnamefont
  {Li}}, \bibinfo {author} {\bibfnamefont {L.}~\bibnamefont {Xu}}, \bibinfo
  {author} {\bibfnamefont {L.}~\bibnamefont {Ding}}, \bibinfo {author}
  {\bibfnamefont {J.}~\bibnamefont {Wang}}, \bibinfo {author} {\bibfnamefont
  {M.}~\bibnamefont {Shen}}, \bibinfo {author} {\bibfnamefont {X.}~\bibnamefont
  {Lu}}, \bibinfo {author} {\bibfnamefont {Z.}~\bibnamefont {Zhu}}, \ and\
  \bibinfo {author} {\bibfnamefont {K.}~\bibnamefont {Behnia}},\ }\href@noop {}
  {\bibfield  {journal} {\bibinfo  {journal} {Phys. Rev. Lett.}\ }\textbf
  {\bibinfo {volume} {119}},\ \bibinfo {pages} {056601} (\bibinfo {year}
  {2017})}\BibitemShut {NoStop}%
\bibitem [{\citenamefont {Kiyohara}\ \emph {et~al.}(2016)\citenamefont
  {Kiyohara}, \citenamefont {Tomita},\ and\ \citenamefont
  {Nakatsuji}}]{Kiyohara_2016}%
  \BibitemOpen
  \bibfield  {author} {\bibinfo {author} {\bibfnamefont {N.}~\bibnamefont
  {Kiyohara}}, \bibinfo {author} {\bibfnamefont {T.}~\bibnamefont {Tomita}}, \
  and\ \bibinfo {author} {\bibfnamefont {S.}~\bibnamefont {Nakatsuji}},\ }\href
  {\doibase 10.1103/PhysRevApplied.5.064009} {\bibfield  {journal} {\bibinfo
  {journal} {Phys. Rev. Applied}\ }\textbf {\bibinfo {volume} {5}},\ \bibinfo
  {pages} {064009} (\bibinfo {year} {2016})}\BibitemShut {NoStop}%
\bibitem [{\citenamefont {Nayak}\ \emph {et~al.}(2016)\citenamefont {Nayak},
  \citenamefont {Fischer}, \citenamefont {Sun}, \citenamefont {Yan},
  \citenamefont {Karel}, \citenamefont {Komarek}, \citenamefont {Shekhar},
  \citenamefont {Kumar}, \citenamefont {Schnelle}, \citenamefont {Kuebler},
  \citenamefont {Felser},\ and\ \citenamefont {Parkin}}]{Nayak2016}%
  \BibitemOpen
  \bibfield  {author} {\bibinfo {author} {\bibfnamefont {A.~K.}\ \bibnamefont
  {Nayak}}, \bibinfo {author} {\bibfnamefont {J.~E.}\ \bibnamefont {Fischer}},
  \bibinfo {author} {\bibfnamefont {Y.}~\bibnamefont {Sun}}, \bibinfo {author}
  {\bibfnamefont {B.}~\bibnamefont {Yan}}, \bibinfo {author} {\bibfnamefont
  {J.}~\bibnamefont {Karel}}, \bibinfo {author} {\bibfnamefont {A.~C.}\
  \bibnamefont {Komarek}}, \bibinfo {author} {\bibfnamefont {C.}~\bibnamefont
  {Shekhar}}, \bibinfo {author} {\bibfnamefont {N.}~\bibnamefont {Kumar}},
  \bibinfo {author} {\bibfnamefont {W.}~\bibnamefont {Schnelle}}, \bibinfo
  {author} {\bibfnamefont {J.}~\bibnamefont {Kuebler}}, \bibinfo {author}
  {\bibfnamefont {C.}~\bibnamefont {Felser}}, \ and\ \bibinfo {author}
  {\bibfnamefont {S.~S.~P.}\ \bibnamefont {Parkin}},\ }\href@noop {} {\bibfield
   {journal} {\bibinfo  {journal} {Sci. Adv.}\ }\textbf {\bibinfo {volume}
  {2}},\ \bibinfo {pages} {e1501870} (\bibinfo {year} {2016})}\BibitemShut
  {NoStop}%
\bibitem [{\citenamefont {Wan}\ \emph {et~al.}(2011)\citenamefont {Wan},
  \citenamefont {Turner}, \citenamefont {Vishwanath},\ and\ \citenamefont
  {Savrasov}}]{Wan2011}%
  \BibitemOpen
  \bibfield  {author} {\bibinfo {author} {\bibfnamefont {X.~G.}\ \bibnamefont
  {Wan}}, \bibinfo {author} {\bibfnamefont {A.~M.}\ \bibnamefont {Turner}},
  \bibinfo {author} {\bibfnamefont {A.}~\bibnamefont {Vishwanath}}, \ and\
  \bibinfo {author} {\bibfnamefont {S.~Y.}\ \bibnamefont {Savrasov}},\
  }\href@noop {} {\bibfield  {journal} {\bibinfo  {journal} {Phys. Rev. B}\
  }\textbf {\bibinfo {volume} {83}},\ \bibinfo {pages} {205101} (\bibinfo
  {year} {2011})}\BibitemShut {NoStop}%
\bibitem [{\citenamefont {Burkov}\ and\ \citenamefont
  {Balents}(2011)}]{Burkov:2011de}%
  \BibitemOpen
  \bibfield  {author} {\bibinfo {author} {\bibfnamefont {A.~A.}\ \bibnamefont
  {Burkov}}\ and\ \bibinfo {author} {\bibfnamefont {L.}~\bibnamefont
  {Balents}},\ }\href@noop {} {\bibfield  {journal} {\bibinfo  {journal} {Phys.
  Rev. Lett.}\ }\textbf {\bibinfo {volume} {107}},\ \bibinfo {pages} {127205}
  (\bibinfo {year} {2011})}\BibitemShut {NoStop}%
\bibitem [{\citenamefont {Shi}\ \emph {et~al.}(2018)\citenamefont {Shi},
  \citenamefont {Muechler}, \citenamefont {Manna}, \citenamefont {Zhang},
  \citenamefont {Koepernik}, \citenamefont {Car}, \citenamefont {Brink},
  \citenamefont {Felser},\ and\ \citenamefont {Sun}}]{Shi_2018}%
  \BibitemOpen
  \bibfield  {author} {\bibinfo {author} {\bibfnamefont {W.}~\bibnamefont
  {Shi}}, \bibinfo {author} {\bibfnamefont {L.}~\bibnamefont {Muechler}},
  \bibinfo {author} {\bibfnamefont {K.}~\bibnamefont {Manna}}, \bibinfo
  {author} {\bibfnamefont {Y.}~\bibnamefont {Zhang}}, \bibinfo {author}
  {\bibfnamefont {K.}~\bibnamefont {Koepernik}}, \bibinfo {author}
  {\bibfnamefont {R.}~\bibnamefont {Car}}, \bibinfo {author} {\bibfnamefont
  {J.~v.~d.}\ \bibnamefont {Brink}}, \bibinfo {author} {\bibfnamefont
  {C.}~\bibnamefont {Felser}}, \ and\ \bibinfo {author} {\bibfnamefont
  {Y.}~\bibnamefont {Sun}},\ }\href@noop {} {\bibfield  {journal} {\bibinfo
  {journal} {arXiv}\ } (\bibinfo {year} {2018})},\ \Eprint
  {http://arxiv.org/abs/arXiv:1801.03273} {arXiv:1801.03273} \BibitemShut
  {NoStop}%
\bibitem [{\citenamefont {Feng}\ \emph {et~al.}(2015)\citenamefont {Feng},
  \citenamefont {Fu}, \citenamefont {Wan}, \citenamefont {Yuan}, \citenamefont
  {Han}, \citenamefont {Quang},\ and\ \citenamefont {Cho}}]{Feng2015}%
  \BibitemOpen
  \bibfield  {author} {\bibinfo {author} {\bibfnamefont {W.}~\bibnamefont
  {Feng}}, \bibinfo {author} {\bibfnamefont {X.}~\bibnamefont {Fu}}, \bibinfo
  {author} {\bibfnamefont {C.}~\bibnamefont {Wan}}, \bibinfo {author}
  {\bibfnamefont {Z.}~\bibnamefont {Yuan}}, \bibinfo {author} {\bibfnamefont
  {X.}~\bibnamefont {Han}}, \bibinfo {author} {\bibfnamefont {N.~V.}\
  \bibnamefont {Quang}}, \ and\ \bibinfo {author} {\bibfnamefont
  {S.}~\bibnamefont {Cho}},\ }\href@noop {} {\bibfield  {journal} {\bibinfo
  {journal} {Phys. Status Solidi RRL}\ }\textbf {\bibinfo {volume} {11}},\
  \bibinfo {pages} {641} (\bibinfo {year} {2015})}\BibitemShut {NoStop}%
\bibitem [{\citenamefont {Kresse}\ and\ \citenamefont
  {Furthm{\"u}ller}(1996)}]{kresse1996}%
  \BibitemOpen
  \bibfield  {author} {\bibinfo {author} {\bibfnamefont {G.}~\bibnamefont
  {Kresse}}\ and\ \bibinfo {author} {\bibfnamefont {J.}~\bibnamefont
  {Furthm{\"u}ller}},\ }\href@noop {} {\bibfield  {journal} {\bibinfo
  {journal} {Phys. Rev. B}\ }\textbf {\bibinfo {volume} {54}},\ \bibinfo
  {pages} {11169} (\bibinfo {year} {1996})}\BibitemShut {NoStop}%
\bibitem [{\citenamefont {Perdew}\ \emph {et~al.}(1996)\citenamefont {Perdew},
  \citenamefont {Burke},\ and\ \citenamefont {Ernzerhof}}]{perdew1996}%
  \BibitemOpen
  \bibfield  {author} {\bibinfo {author} {\bibfnamefont {J.~P.}\ \bibnamefont
  {Perdew}}, \bibinfo {author} {\bibfnamefont {K.}~\bibnamefont {Burke}}, \
  and\ \bibinfo {author} {\bibfnamefont {M.}~\bibnamefont {Ernzerhof}},\
  }\href@noop {} {\bibfield  {journal} {\bibinfo  {journal} {Phys. Rev. Lett.}\
  }\textbf {\bibinfo {volume} {77}},\ \bibinfo {pages} {3865} (\bibinfo {year}
  {1996})}\BibitemShut {NoStop}%
\bibitem [{\citenamefont {Mostofi}\ \emph {et~al.}(2008)\citenamefont
  {Mostofi}, \citenamefont {Yates}, \citenamefont {Lee}, \citenamefont {Souza},
  \citenamefont {Vanderbilt},\ and\ \citenamefont {Marzari}}]{Mostofi2008}%
  \BibitemOpen
  \bibfield  {author} {\bibinfo {author} {\bibfnamefont {A.~A.}\ \bibnamefont
  {Mostofi}}, \bibinfo {author} {\bibfnamefont {J.~R.}\ \bibnamefont {Yates}},
  \bibinfo {author} {\bibfnamefont {Y.-S.}\ \bibnamefont {Lee}}, \bibinfo
  {author} {\bibfnamefont {I.}~\bibnamefont {Souza}}, \bibinfo {author}
  {\bibfnamefont {D.}~\bibnamefont {Vanderbilt}}, \ and\ \bibinfo {author}
  {\bibfnamefont {N.}~\bibnamefont {Marzari}},\ }\href@noop {} {\bibfield
  {journal} {\bibinfo  {journal} {Comput. Phys. Commun.}\ }\textbf {\bibinfo
  {volume} {178}},\ \bibinfo {pages} {685} (\bibinfo {year}
  {2008})}\BibitemShut {NoStop}%
\bibitem [{\citenamefont {Thouless}\ \emph {et~al.}(1982)\citenamefont
  {Thouless}, \citenamefont {Kohmoto}, \citenamefont {Nightingale},\ and\
  \citenamefont {Nijs}}]{Thouless_1982}%
  \BibitemOpen
  \bibfield  {author} {\bibinfo {author} {\bibfnamefont {D.~J.}\ \bibnamefont
  {Thouless}}, \bibinfo {author} {\bibfnamefont {M.}~\bibnamefont {Kohmoto}},
  \bibinfo {author} {\bibfnamefont {M.~P.}\ \bibnamefont {Nightingale}}, \ and\
  \bibinfo {author} {\bibfnamefont {M.}\ \bibnamefont {denNijs}},\ }\href@noop
  {} {\bibfield  {journal} {\bibinfo  {journal} {Phys. Rev. Lett.}\ }\textbf
  {\bibinfo {volume} {49}},\ \bibinfo {pages} {405} (\bibinfo {year}
  {1982})}\BibitemShut {NoStop}%
\bibitem [{\citenamefont {Lu}\ \emph {et~al.}(2015)\citenamefont {Lu},
  \citenamefont {Zhang},\ and\ \citenamefont {Shen}}]{Lu_2015}%
  \BibitemOpen
  \bibfield  {author} {\bibinfo {author} {\bibfnamefont {H.-Z.}\ \bibnamefont
  {Lu}}, \bibinfo {author} {\bibfnamefont {S.-B.}\ \bibnamefont {Zhang}}, \
  and\ \bibinfo {author} {\bibfnamefont {S.-Q.}\ \bibnamefont {Shen}},\
  }\href@noop {} {\bibfield  {journal} {\bibinfo  {journal} {Phys. Rev. B}\
  }\textbf {\bibinfo {volume} {92}},\ \bibinfo {pages} {045203} (\bibinfo
  {year} {2015})}\BibitemShut {NoStop}%
\bibitem [{\citenamefont {Armitage}\ \emph {et~al.}(2018)\citenamefont
  {Armitage}, \citenamefont {Mele},\ and\ \citenamefont
  {Vishwanath}}]{Armitage2018}%
  \BibitemOpen
  \bibfield  {author} {\bibinfo {author} {\bibfnamefont {N.~P.}\ \bibnamefont
  {Armitage}}, \bibinfo {author} {\bibfnamefont {E.~J.}\ \bibnamefont {Mele}},
  \ and\ \bibinfo {author} {\bibfnamefont {A.}~\bibnamefont {Vishwanath}},\
  }\href {https://journals.aps.org/rmp/pdf/10.1103/RevModPhys.90.015001}
  {\bibfield  {journal} {\bibinfo  {journal} {Rev. Mod. Phys.}\ }\textbf
  {\bibinfo {volume} {90}} (\bibinfo {year} {2018})}\BibitemShut {NoStop}%
\bibitem [{\citenamefont {Chadov}\ \emph {et~al.}(2017)\citenamefont {Chadov},
  \citenamefont {Wu}, \citenamefont {Felser},\ and\ \citenamefont
  {Galanakis}}]{Chadov2017}%
  \BibitemOpen
  \bibfield  {author} {\bibinfo {author} {\bibfnamefont {S.}~\bibnamefont
  {Chadov}}, \bibinfo {author} {\bibfnamefont {S.-C.}\ \bibnamefont {Wu}},
  \bibinfo {author} {\bibfnamefont {C.}~\bibnamefont {Felser}}, \ and\ \bibinfo
  {author} {\bibfnamefont {I.}~\bibnamefont {Galanakis}},\ }\href
  {https://journals.aps.org/prb/pdf/10.1103/PhysRevB.96.024435} {\bibfield
  {journal} {\bibinfo  {journal} {Phys. Rev. B}\ }\textbf {\bibinfo {volume}
  {96}},\ \bibinfo {pages} {024435} (\bibinfo {year} {2017})}\BibitemShut {NoStop}%
\end{thebibliography}
\end{document}